   \documentclass[aps,prb,twocolumn,showpacs,superscriptaddress]{revtex4}
   \usepackage{graphicx}
   \usepackage{nicefrac}
   \usepackage{hyperref}
   \usepackage{amsmath}
   \usepackage{amssymb}
   \usepackage{verbatim}
   \usepackage{subfigure}	
   \usepackage{multirow}
   \usepackage{footnote}
   \usepackage{color,soul}
   \newcommand{\qq}{\mathbf{q}}

\begin{document}
   \preprint{}
   
   \title{Confinement effects in ultra-thin ZnO polymorph films: electronic and optical properties}

\author{Lorenzo \surname{Sponza}}
\altaffiliation[Present address: ]{King's College of London, Strand campus, London WC2R 2LS, England, United Kingdom}
\affiliation{CNRS, UMR 7588, Institut des Nanosciences de Paris, F-75005 Paris, France}
\affiliation{Sorbonne Universit\'es, UPMC Univ. Paris 06, UMR 7588, INSP, F-75005 Paris, France}
\author{Jacek \surname{Goniakowski}}
\affiliation{CNRS, UMR 7588, Institut des Nanosciences de Paris, F-75005 Paris, France}
\affiliation{Sorbonne Universit\'es, UPMC Univ. Paris 06, UMR 7588, INSP, F-75005 Paris, France}
\author{Claudine \surname{Noguera}}
\affiliation{CNRS, UMR 7588, Institut des Nanosciences de Paris, F-75005 Paris, France}
\affiliation{Sorbonne Universit\'es, UPMC Univ. Paris 06, UMR 7588, INSP, F-75005 Paris, France}

   \date{\today}

\begin{abstract}
Relying on generalized-gradient and hybrid first-principles simulations, this work provides a complete characterization of the electronic properties of ZnO ultra-thin films, cut along the  Body-Centered-Tetragonal(010), Cubane(100), h-BN(0001),  Zinc-Blende(110), Wurtzite(10$\overline{1}$0) and (0001) orientations. The characteristics of the local densities of states are analyzed in terms of the reduction of the Madelung potential on under-coordinated atoms and surface states/resonances appearing at the top of the VB and bottom of the CB. The gap width in the films is found to be larger than in the corresponding bulks, which is assigned to quantum confinement effects. The components of the high frequency dielectric constant are determined and the absorption spectra of the films are computed. They display specific features just above the absorption threshold due to transitions from or to the surface resonances. This study provides a first understanding of finite size effects on the electronic properties of ZnO thin films and a benchmark which is expected to foster experimental characterization of ultra-thin films via spectroscopic techniques.

\end{abstract}

\maketitle

\section{Introduction}
Zinc oxide is a wide-gap semiconductor \cite{powell+prb1972} of technological importance, widely used for its semiconducting and optical properties. It can be grown under a wide variety of nanoparticle shapes and sizes or as thin films, with many potential applications in the fields of catalysis, gas sensors, photoelectric devices or transparent electronics\cite{ozgur+jap2005, klingshirn+pssb2007, klingshirn+pssb2010}. This richness of applications partly comes from the existence of many structural polymorphs, with rather close formation energies. While wurtzite (WUR) is its structural ground state under normal conditions, ZnO has also been prediceted to adopt zinc-blende (ZB), body-centred tetragonal (BCT), cubane (CUB) or hexagonal Boron Nitride (h-BN) structures, depending upon the conditions of formation\cite{schleife+prb2006, carrasco+prl2007, wang+prb2007, catlow+jcc2008, zhang+cpl2013}. 

In these polymorphs, the zinc and oxygen atoms experience different  environments, especially when some reduction of dimensionality takes place, which may strongly modify the electronic properties. Bulk gap widths and densities of states (DOS) have been scrutinized by first principles simulations \cite{demiroglu+nanoscale2014,zagorac+prb2014} and we have recently disentangled the effects of band narrowing and of electrostatics in the modifications of quasi particle, absorption and electron energy loss spectra\cite{sponza+prb2014}. 

Beyond bulk properties, the possibility of tuning electronic and optical properties through a reduction of dimensionality has fostered advances in the fabrication and structural characterization of ZnO nanostructures \cite{wang+jpcm2004,Shen2009,Bitenc2010,Xu2011,Zhuang2011,Li2011,Han2012} and thin films on various substrates \cite{Chen2000,Hong2002,Smith2003,Koyama2004,Xu2004,Wang2005,Tusche2007,Weirum2010,Kato2010,Xue2010,Schennach2012}, with a range of techniques and under a variety of experimental conditions. In particular in ultra-thin films, theoretical and experimental works indicate  important variations of the 
atomic structure as a function of  thickness\cite{claeyssens+jmc2005,freeman+prl2006,
morgan+prb2009,demiroglu+prl2013}. However, to which extent the electronic properties are affected by these structural changes has not yet been investigated in detail. 

In the present work, we focus on the electronic properties of ZnO thin-films with four monolayer (4ML) thickness, cut along the  BCT(010), CUB(100), h-BN(0001),  ZB(110), WUR(10$\overline{1}$0) and (0001) orientations. We analyze the gap widths, the layer-projected density of states  and the optical properties. We are able to highlight the role of surface under-coordinated atoms in determining the largest modifications of these properties with respect to the corresponding bulk structures.

The structure of the paper is the following. After a description of the numerical approach  (Section \ref{sec:method}), we stress the main structural characteristics which will be of importance for the understanding of the electronic properties (Section \ref{sec:structure}). We then analyze the DOS characteristics, with special emphasis on electrostatic  effects, and we discuss the gap widths (Section \ref{sec:electronic}). Optical dielectric functions and absorption spectra are the subject of Section \ref{sec:optics}, before the conclusion.

\section{Method}\label{sec:method}

All ground state calculations are performed within the framework of the Density Functional
Theory (DFT), using the projector
augmented wave method\cite{blochl+prb1994,VASP_paw}, and a plane wave basis set, as implemented in VASP\cite{VASP_code}. Valence electrons are $2s$ and $2p$ for oxygen, $3d$ and $4s$ for zinc. The Perdew, Burke and Ernzerhof (PBE) generalized-gradient approximation\cite{perdew+prl1996}
is systematically used for structural optimization, while electronic properties are computed with the range-separated hybrid HSE03 exchange-correlation functional \cite{anisimov+jpcm1997}. The use of the HSE03 functional is meant to improve the PBE underestimation of the band gap in semiconductors and insulators and the excessive delocalization of $d$ electrons. It is especially useful for the simulation of compounds with full or empty $d$ shells, to the description of which the DFT+U approach is less efficient.

Slabs of 4ML \cite{note1} have been designed for each structure and orientation, with approximately 12 \AA\ of empty space  to prevent spurious interactions between replicated slabs.  Dipole correction is applied to slabs with two non-equivalent terminations.
The energy cutoff is 500 eV and the k-point grids, centered at $\Gamma$, 
	used to sample the Brillouin zone of the ($1 \times 1$) surface cell, are 
	5x7x1 for BCT(010),
	4x4x1 for CUB(100),
	8x8x1 for h-BN(0001),
	8x6x1 for WUR(10$\bar{1}$0), 
	8x8x1 for ZB(110) and 
	6x6x1 for WUR(0001).
Cell optimization is stopped when all forces get lower than 0.01 eV/\AA\ 
	and in-plane components of the 	stress tensor below 0.01 eV/\AA$^3$.  
This setting leads to converged values of the cell-parameters within 0.01 \AA\, and of the total energies within 0.01 eV per formula unit. Only the ideal BCT  structure ($a=b$) is considered.

Optical properties are computed within the linear response theory, 
in which the response to a perturbing field is described by the complex dielectric function $\epsilon(\qq,\omega) = \epsilon_1(\qq,\omega) + i \epsilon_2(\qq,\omega)$ ($\omega$  the energy of the perturbation  
and $\qq$ the exchanged momentum which tends to zero for interaction with light).
In the random phase approximation\cite{onida+rmp2002} (RPA), 
	and under the assumption of an homogeneous medium (neglect of local fields), 
	the dielectric function $\epsilon(\qq,\omega) = 1 - 4\pi\chi^0(\qq,\omega) / |\qq|^2$
	is expressed in terms of the macroscopic component of the independent particle polarizability $\chi^0(\qq,\omega)$, that we compute as the weighted sum over all transitions\cite{adler+pr1962,wiser+pr1963} from occupied to empty (HSE03) Kohn-Sham states, thus neglecting excitonic effects.
Within such approximations, the absorption spectrum is given by $A(\omega)=\lim_{\qq\rightarrow 0} \epsilon_2(\qq,\omega)$. Depending on the Cartesian direction along which the limit is taken, the spectra along $x$, $y$ or $z$ direction are computed ($z$ perpendicular to the surface). The dielectric constant $\lim_{\mathbf{q} \rightarrow 0}\epsilon_1({\mathbf{q}},0)$ is computed via Kramers Kronig relations from the imaginary part $\epsilon_2(\qq,\omega)$. 
The sum over all transitions has been cut at 30 eV and  finer k-point grids (all centered in $\Gamma$) are used:
12x9x1 for WUR(10$\bar{1}$0), 9x12x1 for BCT(010), 9x9x1 for CUB(100) and 12x12x1 for the remaining three structures.

\section{Structural properties}\label{sec:structure}

\begin{figure}
 	\centering
 	\includegraphics[width=0.5\textwidth]{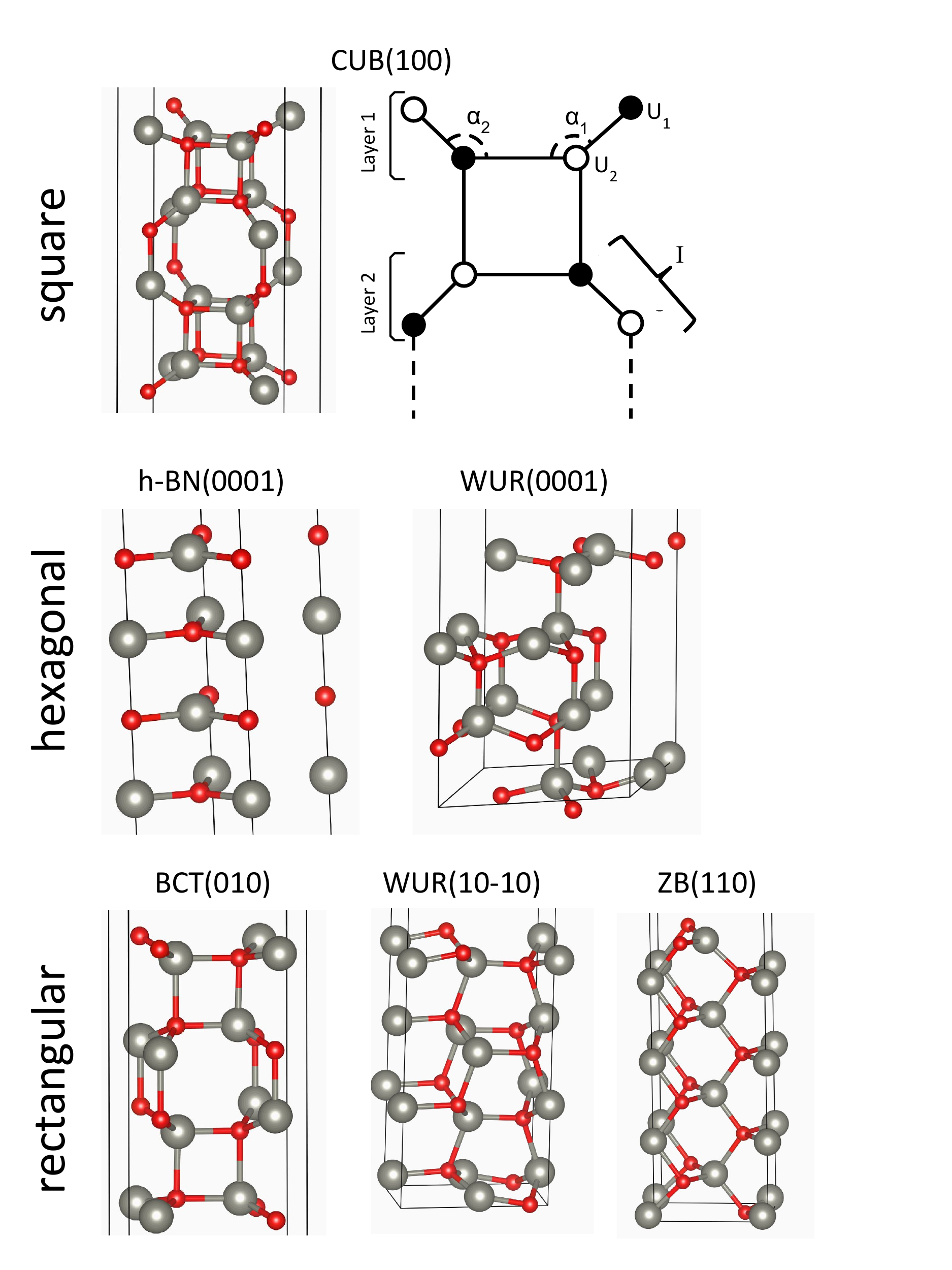}
 	\caption{(Colors online:) Unit cells of 4ML ZnO thin-films. 
 		Zn and O atoms are represented by big (grey) and small (red) balls, respectively.
 		For CUB(100) a sketch of the structure clarifies the convention for the sites and the layers. }
 	\label{fig:structures}
 \end{figure}

This section is devoted to a description of the main structural characteristics of the  4ML films, with a special emphasis on those which will be of importance for understanding the electronic properties. For each polymorph, we have considered the film orientation which has the lowest energy, i.e. BCT(010), CUB(100), h-BN(0001), ZB(110) and WUR(10$\overline{1}$0)\cite{Meyer2003,claeyssens+jmc2005}. We have also considered the WUR(0001) polar orientation, with a reconstructed ($2\times 2$) surface configuration in which one oxygen (resp. zinc) atom upon 4 is removed on the oxygen  (resp. zinc) termination. This is the conventional method to reduce the build-up of a macroscopic dipole. The 2D unit cells of these 4ML films are sketched in Figure~\ref{fig:structures}. The surface unit cells display different symmetries: square  for CUB(100), hexagonal  in the case of h-BN(0001) and WUR(0001), and
rectangular  for BCT(010), WUR(10$\bar{1}$0) and ZB(110). At the film surfaces, some atoms are under-coordinated. We label them $U_1$ and $U_2$ according to whether they are directly in contact with vacuum or immediately sub-surface. Internal atoms are labeled $I$, as sketched in Figure~\ref{fig:structures}. In the h-BN(0001) films, $U_2$ atoms do not exist. In the case of WUR(0001), there are two types of sub-surface atoms with different local environments: three atoms (labeled $U_3$  in the following) are adjacent to the surface vacancy while the fourth one (labeled $U_2$) is not.

Structural data, compared to reference bulk calculations\cite{sponza+prb2014}, 
are reported in Table~\ref{tab:structural_data}, as well as the HSE03 film formation energy per surface area, defined as:
\begin{equation}
 E_f = \frac{E_{film} - nE_{bulk}^{WUR}}{S} 
 \label{eq:surfen}
\end{equation}
In this expression, $n$ is the number of formula units in the thin-film cell, 
	$E_{film}$ is the total energy of the film per unit cell,
	$E_{bulk}^{WUR}$ is the total energy per formula unit of the bulk wurtzite  structure, both calculated at the HSE03 level on top of the PBE structural ground state,
and $S$ is the total surface of the film cell including upper and lower surfaces.
Our results both on the structural parameters and the energetics compare well with similar \textit{ab-initio} calculations found in the literature\cite{demiroglu+prl2013,claeyssens+jmc2005,freeman+prl2006,morgan+prb2009,goniakowski+prl2007}.

\begin{table}[h!]

	\centering
    \caption{Structural and energetic properties of ZnO 4ML thin-films: surface unit cell parameters (in-plane $a$,$b$), film thickness $h$, surface layer average angle $\langle \alpha \rangle= (\alpha_1+\alpha_2)/2$, as shown in Figure \ref{fig:structures},  dangling bond angle $\theta$ (see text), and formation energy $E_f$ per surface area. Structural quantities are compared to corresponding bulk values\cite{sponza+prb2014} (in parenthesis).}
	\begin{tabular}{ l l}
	\hline
	\multicolumn{2}{l}{WUR(10$\bar{1}$0): rectangular 2D unit cell} \\
		 \quad 		 $a$ (\AA) & 3.32 (+1.2\%  wrt bulk) \\
		 \quad 		 $b$ (\AA) & 5.34 (+0.8\%  wrt bulk) \\
		 \quad 		 $h$ (\AA) & 9.39 (-1.0\%  wrt bulk)  \\
		 \quad		 $\langle \alpha \rangle$ ($^\circ$) & 112.7 (+3.7\% wrt bulk)\\
		 \quad       $\theta$ ($^\circ$)&10\\
		 \quad 		 $E_f$ (eV/\AA$^2$)\;\; & 0.058 \\ 
	\multicolumn{2}{l}{BCT(010): rectangular 2D unit cell} \\
		 \quad 		 $a$ (\AA) & 5.70 (+1.4\%  wrt bulk) \\
		 \quad 		 $b$ (\AA) & 3.31 (+0.9\%  wrt bulk) \\
		 \quad 		 $h$ (\AA) & 8.69 (-5.4\%  wrt bulk)  \\
 		 \quad		 $\langle \alpha \rangle$ ($^\circ$) & 117.9 (+4.3\% wrt bulk)\\
 		 \quad       $\theta$ ($^\circ$)&6\\
		 \quad 		 $E_f$ (eV/\AA$^2$)\;\; & 0.049 \\ 
	\multicolumn{2}{l}{ZB(110): rectangular 2D unit cell} \\
		 \quad 		 $a$ (\AA) & 3.31 (+1.2\%  wrt bulk) \\
		 \quad 		 $b$ (\AA) & 4.60 (-0.4\%  wrt bulk) \\
		 \quad 		 $h$ (\AA) & 11.41 (-0.3\%  wrt bulk)  \\
 		 \quad		 $\langle \alpha \rangle$ ($^\circ$) & 111.8 (+2.1\% wrt bulk)\\
 		  \quad      $\theta$ ($^\circ$)&28\\
		 \quad      $E_f$ (eV/\AA$^2$)\;\; & 0.067 \\		 
	\multicolumn{2}{l}{CUB(100): square 2D unit cell} \\
		 \quad 		 $a=b$ (\AA) & 6.34 (+1.0\%  wrt bulk) \\
		 \quad 		 $h$ (\AA) & 10.05 (-4.3\%  wrt bulk)  \\
 		 \quad		 $\langle \alpha \rangle$ ($^\circ$) & 129.8 (+3.6\% wrt bulk)\\
 		 \quad       $\theta$ ($^\circ$)&12\\
		 \quad 		 $E_f$ (eV/\AA$^2$)\;\; & 0.078  \\ 		 
	\multicolumn{2}{l}{h-BN(0001): hexagonal 2D unit cell} \\
		 \quad 		 $a=b$ (\AA) & 3.39 (-2.6\%  wrt bulk) \\
		 \quad 		 $h$ (\AA) & 7.11 (+4.8\%  wrt bulk)  \\
		 \quad       $\theta$ ($^\circ$)&0\\
		 \quad 		 $E_f$ (eV/\AA$^2$)\;\; & 0.049 \\ 
	\multicolumn{2}{l}{WUR(0001): hexagonal 2D unit cell} \\	
		 \quad 		 $a=b$ (\AA) & 6.62 (+0.9\%  wrt bulk) \\		 
		 \quad 		 $h$ (\AA) & 8.11 (-5.5\%  wrt bulk)  \\ 
		 \quad       $\theta$ ($^\circ$)&0\\  
		 \quad 		 $E_f$ (eV/\AA$^2$)\;\; & 0.086  \\
		\hline
	\end{tabular}
	\label{tab:structural_data}
\end{table}

A feature shared by all films except h-BN(0001) is the expansion of the lateral lattice parameters, with respect to bulk. This effect is caused by a flattening of the surface layers, evidenced by an increase in the average surface bond angle $\langle \alpha \rangle$. This is the so-called rotational relaxation mechanism, well-known at the surface of semiconductors \cite{Chadi1978,Chadi1979}.
It induces a contraction of the structure in the perpendicular direction, to preserve the atomic volumes. However, in the present case, it does not preclude some simultaneous bond contraction. In the h-BN(0001) film, since the layers are already flat, surface bond breaking only induces a contraction of the in-plane parameters, which leads to an expansion in the perpendicular direction.  

More detailed analysis  (Table \ref{tab:dos_details} in Section \ref{sec:electronic}) shows that, in all structures, under-coordinated atoms $U_1$ or $U_3$ have lost one first neighbor and between 3 and 5 second neighbors, while atoms $U_2$ have a complete first coordination shell and a second coordination shell reduced by 1 to 3 units. The average Zn-O bond-lengths around $U_1$ and $U_3$ atoms are reduced by approximately 0.1 \AA\  ({see Table I in Supplemental Material\cite{supp}}). In all cases, $I$ atoms have complete first and second coordination shells, and the bond relaxation around them is quasi-negligible in average.
	
Finally, in Table \ref{tab:structural_data}, we give the angle $\theta$ between the $ab$ plane and the plane which contains the three first neighbors of U$_1$ atoms. This angle measures the projection of the dangling bond located on U$_1$ atoms on the $z$ axis perpendicular to the surface, and thus characterizes the mixing of $p_z$ with $p_x,p_y$ orbitals in the surface dangling bonds.

\section{Electronic structure}\label{sec:electronic}

In this section, we discuss the electronic characteristics, obtained by using the hybrid HSE03 exchange-correlation functional on top of the PBE structural ground state of the films. The relevant characteristics include the main DOS structures in the valence (VB) and conduction (CB) bands, and the gap width. We relate their modifications with respect to their respective  bulks to changes in the local environment of the surface atoms.

\begin{figure}[h]
	\centering
\includegraphics[width=0.55\textwidth]{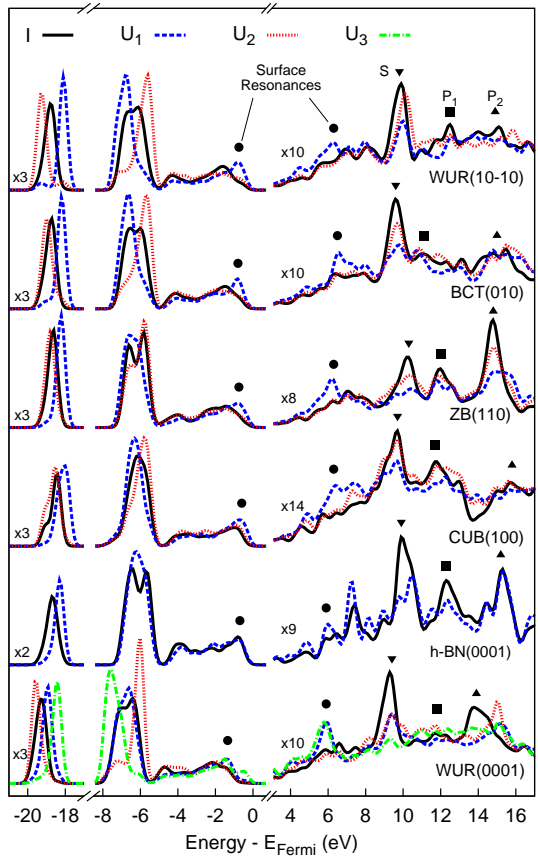}
\vspace{-1cm}\caption{(Color online) Local densities of states in ZnO 4ML films on I, $U_1$, $U_2$ and $U_3$ (in the case of WUR(0001)) atoms, represented as solid black,  dashed blue,  dotted red, and dashed-dotted green lines, respectively.} 
\label{fig:dos}
\end{figure}

\subsection{Characteristics of the local densities of states}\label{sec:electronic-DOS}

Figure~\ref{fig:dos} displays the local densities of states (LDOS) of the six films, projected on internal layers (I atoms), surface atoms of type U$_1$ and sub-surface atoms of type U$_2$ or $U_3$. 

The overall DOS shape is similar for all polymorphs. At the very bottom of the VB, two peaks with nearly O ${2s}$  and Zn $3d$ character, respectively, are found, lower in energy than the  oxygen band, formed by the O$_1$ peak ($\sim -5$ eV) due to Zn $4s$-O $2p$ bonding states, 
	and  the O$_2$ region (from Fermi level to about -4 eV), associated to Zn $3d$-O $2p$ anti-bonding states and O-O hybridization. In the conduction band, beyond the extended onset region, three zones can be identified: 
	the $S$ zone, mainly formed by O $2p$ - Zn $4s$ anti-bonding states,
	the $P_1$ zone, formed by states with mixed Zn $p$ and Zn $s$ character,
	and the $P_2$ zone at higher energy, with predominant Zn $p$ orbital component.
The average positions of the most noticeable structures are reported in Table ~\ref{tab:dos_peaks}, together with a comparison of gap widths $E_g$ between the bulks and the films.

\begin{table*} 
	\centering
	\caption{Mean positions of the main LDOS structures on the internal $I$ layers (see text) with respect to the Fermi level (in eV); bulk and film gap widths $E_g$ and their difference $\delta E_g=E_g$(film)-$E_g$(bulk) (in eV). Note that for h-BN the bulk gap value recorded here is the minimal gap; the direct one is equal to 2.42 eV}
	\begin{tabular}{| l | c c c c| c c c | c c c | }
	\hline
	& O  $2s$& Zn  $3d$& O$_1$& O$_2$ &  $S$ & $P_1$ & $P_2$& $E_g$(bulk)  & $E_g$(film) & $\delta E_g$ \\
	\hline
WUR(10$\bar{1}$0) &  -18.83 & -6.36 & -4.14 & -1.86 & 9.8 & 12.5  & 14.9& 2.20&2.25&+0.05 \\
BCT(010)&  -18.81 & -6.30 & -3.94 & -1.72 &       9.6  & 11.1 & 15.0&2.20&2.37&+0.17 \\
ZB(110) &   -18.70 & -6.19 & -4.21 & -1.83  & 10.3 & 12.0 & 14.8&2.09 &2.31&+0.22  \\
CUB(100)& -18.56 & -6.18  & -3.64 & -1.48 & 9.7  & 11.7  & 15.8&2.68 &2.71&+0.03\\  
h-BN(0001) & -18.72 & -6.15 & -3.92  & -1.67 & 9.9  & 12.3  & 15.2&2.31&2.66&+0.35\\
WUR(0001)& -19.21 & -6.63 & -4.61  & -2.35 & 9.4  & 11.8  & 13.9&2.20	&1.62&-0.58\\
	\hline	
	\end{tabular}
	\label{tab:dos_peaks}
	\end{table*}

There is a  close resemblance between the LDOS shapes that are found on the internal I atoms and those in the bulk, displayed in our previous work \cite{sponza+prb2014}. The tiny differences, in particular the splitting of the Zn d-state due to the crystal field which is somewhat blurred in some films, result from the long range electrostatic and/or covalent interactions existing in ZnO, which are partly cut in the films.

As far as LDOS on surface atoms $U_1$, $U_2$ or $U_3$ are concerned, it is difficult to find systematic characteristics of the S, P$_1$ or P$_2$ peaks in the CB, likely because of the delocalized nature of the orbitals involved, except in the two WUR films for which the S and P$_2$ peaks of the $U_1$ and $U_2$ LDOS are shifted towards higher energies. Close to the VB and CB band edges, there is a clear enhancement of the surface LDOS on under-coordinated atoms, but, considering the orbital overlap between surface and sub-surface atoms and the small width of the films, it is difficult to discriminate between actual surface states and surface resonances. In the CB, for all polymorphs, surface resonances mainly localized on $U_1$ and $U_2$ atoms  are present from approximately 2 to 4 eV above the CB minimum. The CB minimum itself involves orbitals on all atoms of the films, except in the WUR(0001) film, where there is a well-defined surface state on $U_3$ atoms. 

In the VB, surface resonances with strong $U_1$ and $U_2$ oxygen $p_z$ character and to a lesser extent  I oxygen orbitals are  present in all polymorphs  in an energy range from 0.2 to $\approx$ 1 eV below the top of the VB. At the extreme top of the VB  true surface states  with $U_1/U_3$ and $U_2$ oxygen $p_x$ and $p_y$ character are present, with a degree of hybridization with $p_z$ orbitals which increases in the series: h-BN(0001), WUR(0001), BCT(010), WUR(10$\bar{1}$0), CUB(100), ZB(110) ({see Figure 1 in Supplemental Material\cite{supp}}). As expected, the evolution of the degree of hybridization follows the variations of the angle $\theta$ between the U$_1$ dangling bond orientation and the direction $z$ perpendicular to the surface which increases in the series (Table \ref{tab:structural_data}).
The relationship between these VB LDOS characteristics, the absorption spectra and the optical dielectric function will be discussed in Section \ref{sec:optics}.

More remarkable are the differences between internal and surface LDOS involving  localized O$2s$ and Zn$3d$ orbitals in the valence band. Figure \ref{fig:dos} evidences a systematic shift towards higher energies (blue-shift) for the former and towards lower energies (red-shift) for the latter on the $U_1$ or $U_3$ atoms, while the opposite is true on $U_2$ atoms. These shifts are given in Table \ref{tab:dos_details} and are related to the modifications of the electrostatic potential on under-coordinated atoms, to be discussed below.

\subsection{Electrostatic effects}
LDOS characteristics, especially as far as localized states are concerned, may be understood by considering the electrostatic potentials\cite{noteelectrostatic} $V_O$ and $V_{Zn}$ acting on the various oxygen and zinc atoms of the films. 
They include long range contributions, but their variations among nonequivalent atoms are mainly related to local environments (i.e. to the variations $\delta N_1$ and $\delta N_2$ of the numbers of first and second neighbors). Table \ref{tab:dos_details} records the differences $\delta V_O$ and $\delta V_{Zn}$ of electrostatic potentials between under-coordinated surface atoms and fully-coordinated internal atoms. 

\begin{table}[h]
	\centering
	\caption{Shifts of the average peak positions (in eV), 
		variations of electrostatic potentials $\delta V_O$ and $\delta V_{Zn}$ on oxygen and Zn atoms (in V) and of  coordination numbers $N_1$ and $N_2$ on under-coordinated $U_1$, $U_2$ and $U_3$ atoms, with respect to internal atoms I.}
	\begin{tabular}{| l c c c   | c c | c c |}
	\hline
	&   O $2s$ \,   & \,   Zn $3d$  \, &    \, O $2p$\, & $\delta V_O$\;  &$\delta V_{Zn}$\;\;  & $\delta N_1$ & $\delta N_2$\\
	\hline
&\multicolumn{2}{l}{\ WUR(10$\bar{1}$0):} & & & && \\
	\qquad U$_1$-I &  +0.64  & -0.37 & +0.53 & -0.76 &+0.47 & -1 & -4\\ 
	\qquad U$_2$-I & -0.33 & +0.42 & -0.09  & +0.37 &-0.54 & 0 & -2\\
&\multicolumn{2}{l}{\ BCT(010):}   & & && &\\
	\qquad U$_1$-I & +0.55   & -0.27 & +0.43 & -0.60 &+0.36 & -1 & -3  \\ 
	\qquad U$_2$-I & -0.11 & +0.33 & -0.01   & +0.19 &-0.39 & 0   & -2\\ 
&\multicolumn{2}{l}{\ ZB(110):}   & & & && \\	
	\qquad U$_1$-I &  +0.43  &  -0.20 & +0.42  & -0.51 &+0.25 & -1 & -5 \\
	\qquad U$_2$-I &  -0.09   & +0.07 & -0.01   & +0.08 &-0.10 & 0  &  -1\\ 
&\multicolumn{2}{l}{\ CUB(100):}   & & & & &\\
	\qquad U$_1$-I & +0.42  & -0.15  & +0.38 & -0.45 &+0.20 & -1 & -3 \\	
	\qquad U$_2$-I & -0.08 & +0.15 & +0.01   & +0.09 &-0.17 & 0 & -1\\ 
&\multicolumn{2}{l}{\ h-BN(0001):}   & & & &&\\
	\qquad U$_1$-I     & +0.37   & -0.02 & +0.18  & -0.18 &+0.10 & -1 & -3\\ 
&\multicolumn{2}{l}{\ WUR(0001):}   & & & && \\	
	\qquad U$_1$-I   & 0.35   & -0.05 & +0.41 & -0.40& +0.08    & -1  & -5 \\  
	\qquad U$_{3}$-I & +0.78  & -0.63 & +0.66 &  -0.92 &+0.75 & -1  & -3 \\ 
	\qquad U$_{2}$-I & -0.18  & +0.41 & -0.10 &  +0.23 &-0.58 & 0   & -3  \\  
 	\hline	
	\end{tabular}
	\label{tab:dos_details}
	\end{table}

On $U_1$/$U_3$ atoms, the negative sign of $\delta V_O$ and the positive sign of $\delta V_{Zn}$ indicate a decrease (in absolute value) of the electrostatic potential, as expected from the loss of one first neighbor. The effect is modulated by longer range interactions and by the length of the broken bond. For example, in the h-BN film, which has a quasi layered structure, the bonds which are broken at the surface are very long and do not induce strong modifications of electrostatic potential. Consequently, the latter is mostly determined by the in-plane arrangement, which does not change from plane to plane. In WUR(10$\overline{1}$0) and (0001), BCT(010), ZB(110) and CUB(100) films, $\delta V_O$ and $\delta V_{Zn}$ are quite noticeable. WUR(0001) film displays the largest potential variations, which we will discuss separately. Conversely, as far as  $U_2$ atoms are concerned,  the signs of $\delta V_O$ and $\delta V_{Zn}$  are opposite to those on $U_1$ and $U_3$, consistently with no loss of first neighbors (of opposite charge) and a decrease in the number of second neighbors (of same charge).

\begin{figure}[h]
	\includegraphics[width=0.50\textwidth]{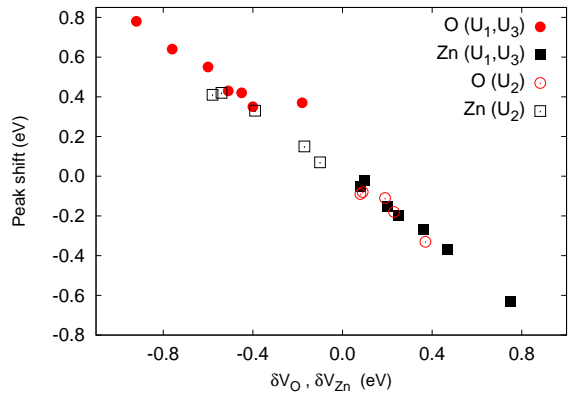}
	\caption{(Color online).
	Linear correlation between the positions with respect to I of the Zn3$d$  and O $2s$ LDOS peaks localized on atoms U$_1$, $U_2$ and U$_3$  and the electrostatic potential differences $\delta V_{O}$ or $\delta V_{Zn}$ for all polymorph films (see text).}
	\label{fig:zn3dVSdv}
\end{figure}

Within a Hartree or Hartree-Fock approximation, the diagonal matrix elements of the Hamiltonian/Fock operator on an atomic orbital basis set have a contribution $-V_i$ due to the electrostatic potential\cite{Nogueralivre,Noguera2001}. Applied to ZnO films, the decrease of $|\delta V|$ on  $U_1$ and $U_3$ atoms thus pushes their O $2s$ levels towards higher energies and their Zn $3d$ levels towards lower energies. Conversely, the increase of $|\delta V|$ on $U_2$ atoms pushes their O $2s$ levels and their Zn $3d$ levels in the opposite direction. Figure \ref{fig:zn3dVSdv} shows that the correlation is quantitatively obeyed, for all under-coordinated atoms.

The reduction of electrostatic potential on $U_1/U_3$ atoms is also responsible for the presence of strong surface states/resonances in the top part of the VB and the bottom part of the CB. However, the hybridized character of these states makes the correlation between their positions and the reduction of electrostatic potential less quantitative than for the more localized O $2s$ and Zn $3d$ states, and the precise surface geometry also plays a role. At the very bottom of the CB  where the states are more diffuse, no actual surface state exists, only resonances with a substantial admixing with I atom orbitals, despite the values of $\delta V_O$ and $\delta V_{Zn}$. More localized DOS structures corresponding to surface resonances may be found from approximately 2 to 4 eV above the CB minimum (Figure \ref{fig:dos}).

At  the very top of the VB, in WUR(10$\overline{1}$0), WUR(0001), BCT(010), and CUB(100) films, $\delta V_O$ and $\delta V_{Zn}$ are sufficiently large to produce surface states and surface resonances. Highest in energy are weak oxygen-oxygen anti-bonding states mainly involving $p_x$ and $p_y$ orbitals within the surface layer, while a few tenths of eV below are surface resonances of oxygen $p_z$ dangling bond character. In ZB(110) films, little oxygen-oxygen bonding exists within the surface layer, so that the states at the very top of the VB have a resonance, not a surface state, character.  As well recognized at the surface of ZB semi-conductors, the buckling of the surface dimers which  pushes oxygens outwards and tends to transform the sp$_3$ bonding into $sp_2$ hybridization pushes back the dangling bond surface state into the VB. In the h-BN(0001) film, all layers being nearly identical from an electronic point of view due to the extremely weak interlayer coupling (reflected in the weak $\delta V_O$ and $\delta V_{Zn}$ values), VB states display a mixed character between U$_1$ and I orbitals. 
Finally, the largest electrostatic potential variations are found at the surface of WUR(0001) on $U_3$ atoms. They induce the presence of surface states with $p_x$, $p_y$ character, not only at the top of the VB but also at the bottom of the CB. 

As shown in Table \ref{tab:dos_details}, in WUR(0001), there is a surprisingly large difference in $\delta V_O$ and $\delta V_{Zn}$ values between $U_1$ and $U_3$ atoms, while their local environments present the same reduction of first neighbors and a not so much different reduction of second neighbors. Remembering that atoms U$_1$ and U$_3$ are of same chemical nature and are located at opposite film terminations, we assign the difference to a residual electrostatic dipole, due to an incomplete compensation of polarity. Indeed, at the WUR(0001) semi-infinite polar surface, it is usually considered that the removal of one fourth of surface ions does heal polarity because the ratio of successive interlayer distances $R_1/(R_1+R_2)$ is close to 1/4. However this statement neglects the fact that  the $R_1/(R_1+R_2)$ ratio is not exactly equal to 1/4. It also neglects the  polarization of electronic origin, which exists in the non-centro-symmetric wurtzite structure. Moreover, in the 4ML film, finite size effects have to be taken into account, which modify the criterion of polarity compensation \cite{Noguera2008,Noguera2013}. This interpretation is further confirmed by an analogous simulation of a 4ML ZB(111) film. Since along the ZB(111) polar orientation  $R_1/(R_1+R_2)=1/4$ and since there is no spontaneous polarization in the ZB centro-symmetric structure,  the removal of one fourth of surface ions exactly heals polarity at its semi-infinite (111) surface. In ZB(111) thin films, thus only finite size effects may produce a residual dipole. We indeed find that the difference in $\delta V_O$ and $\delta V_{Zn}$ values between U$_1$ and U$_3$ atoms is strongly reduced compared to that in WUR(0001) (-0.02 V and +0.17 V on oxygen and zinc atoms, respectively, to be compared to -0.52 V and +0.67 V).

\subsection{Band gaps}
As shown in Table \ref{tab:dos_peaks}, the HOMO-LUMO gaps of the films are systematically larger than their bulk counterparts, except in the WUR(0001) film. Several effects  affect the gap widths. The first one, which has been invoked to explain gap variations in ZnO bulk polymorphs \cite{demiroglu+nanoscale2014,sponza+prb2014}, is a gap enlargement due to the decrease of surface band widths. Indeed, surface atoms have less second neighbors than internal atoms, leading to a narrowing of the O$2p$-O$2p$ band width at the top of the valence band and of the Zn$4s$-Zn$4s$ band width at the bottom of the conduction band. However, this surface effect does not modify the minimal gap of the film, which is found in the LDOS of internal atoms. A second effect is a gap reduction due to the decrease of the electrostatic potential on surface atoms, as discussed above. Surface states  present at the top of VB or bottom of CB induce a gap reduction. We have  previously seen that the effect is especially strong in the WUR(0001) film but also exists, although to a lesser extent, in WUR(10$\overline{1}$0), BCT(010), and CUB(100) films. However,  Table \ref{tab:dos_peaks} shows that, in all cases except WUR(0001), this electrostatic-driven gap reduction is not dominant. The gaps in the films are larger than in the bulks, which requires larger gaps in the  LDOS of I atoms than in the bulks. Quantum confinement, i.e. the quantification of  states propagating perpendicular to the film surfaces, can induce such opening of the gap. This effect, which asymptotically decreases as the films become thicker, is expected to be particularly strong in such ultra-thin films and to prevail over the gap narrowing due to electrostatic effects, except when the latter is exceptionally strong as in WUR(0001).

\section{Optical properties}\label{sec:optics}
On the basis of the electronic structure characteristics discussed in the preceding section, we now analyze the optical properties of the ZnO  films, first focusing on the optical dielectric constant and then on the absorption spectra.

\subsection{Optical dielectric function}
The values of the  high frequency bulk dielectric constant $\epsilon _\infty$, i.e. the values of the  dielectric constant in the limit of infinite phonon  and zero electronic frequencies, are gathered in Table \ref{tab:cstopt}, for the various ZnO polymorphs. In anisotropic bulks (WUR, BCT and h-BN), the ordinary (OC) and extraordinary (EC) components are given, associated to a momentum transfer  parallel and perpendicular to the {\it ab} plane, respectively. 

\begin{table}[h]
	\centering
	\caption{High frequency dielectric function, in the bulk (left part of the table) and in the films (right part), with reference to the mean electronic density $n$ in the bulk (in \AA$^{-3}$). OC and EC denote the ordinary and extraordinary components, respectively. In the films, $x$ and $y$ refer to directions parallel to the surface, while $z$ is perpendicular to it.}
	\begin{tabular}{| l | c c c || c c c  c | }
	\hline
	&  OC& EC & $n$&  & x &y & z \\
	\hline
WUR &3.50    & 3.54  & 1.54  & (10$\bar{1}$0) & 4.04  &4.12   & 3.72 \\
 &    &   &   & (0001) & 4.22  &4.22   & 3.97 \\
BCT & 3.35  &3.44  & 1.47  & (010) &4.09   &  4.09 & 3.69 \\
ZB & 3.59   &3.59   & 1.54  & (110) &3.96   &3.88   &3.72  \\
CUB & 2.92   &2.92   & 1.23  & (100) & 3.46  &3.46   &3.19  \\
h-BN & 3.58   & 3.68  & 1.60  & (0001) & 4.53  & 4.53  & 4.09 \\
	\hline	
	\end{tabular}
	\label{tab:cstopt}
	\end{table}

In bulk WUR, the values found are in agreement with previous theoretical estimations at the same level of theory \cite{Wrobel2009,Schleife2009,Gori2010}, and close to the experimental value of 3.7 \cite{Martienssen2005}, a typical value for a semiconductor of mixed ionic and covalent character. Among the polymorphs, the variations of $\epsilon _\infty$ qualitatively correlate  with the mean valence electronic density $n$, as expected from simple models of screening  \cite{Nogueralivre} $\epsilon _\infty\approx 1+4\pi ne^2/(mE_g^2)$.  

The evaluation of the high frequency  dielectric constants $\epsilon_\infty$ of the films is less straightforward. In a first step, the dielectric constant $\epsilon_{sc}(L)$ of the simulated supercell, which includes the film (of thickness $h$) and a large vacuum thickness $L$, is calculated for several values of $L$ and shown to depend  on $L$ according to the following law:
\begin{equation}
\epsilon_{sc}(L)=\frac{h*\epsilon_\infty+L*\epsilon_{vac}}{L+h}
\end{equation}
in which $\epsilon_{vac}=1$ is the vacuum dielectric constant. This law, which is an extrapolation of 
the expression derived in Ref. \onlinecite{Cudazzo2011} for the dielectric screening in two-dimensional insulators, assumes that the electronic polarizability of a complex medium can be estimated from an  average of the polarizabilities of its various parts weighted by their respective volumes. From the extrapolation at $L\to 0$ of the linear plot of $(L+h)\epsilon_{sc}(L)/h$, the film dielectric function $\epsilon_\infty$ can be obtained. Its three components associated to  momentum transfers along $x$, $y$ and $z$ are given in Table \ref{tab:cstopt}.

As a general statement, the components of the film dielectric functions are larger than in the bulk by c.a. 20\%, but still correlate with the mean electronic density. The anisotropy between the $x$ and $y$ components is nonexistent (WUR(0001), CUB(100), h-BN(0001)) or extremely weak (WUR(10$\bar{1}$0), BCT(010)\cite{anisotropy}, ZB(110)), while the $z$ component is systematically smaller than the $x$ and $y$ ones. The overall anisotropy is thus larger in thin films than in the bulks.

\begin{figure}[h] 
\begin{center}
	\includegraphics[height=0.66\textheight]{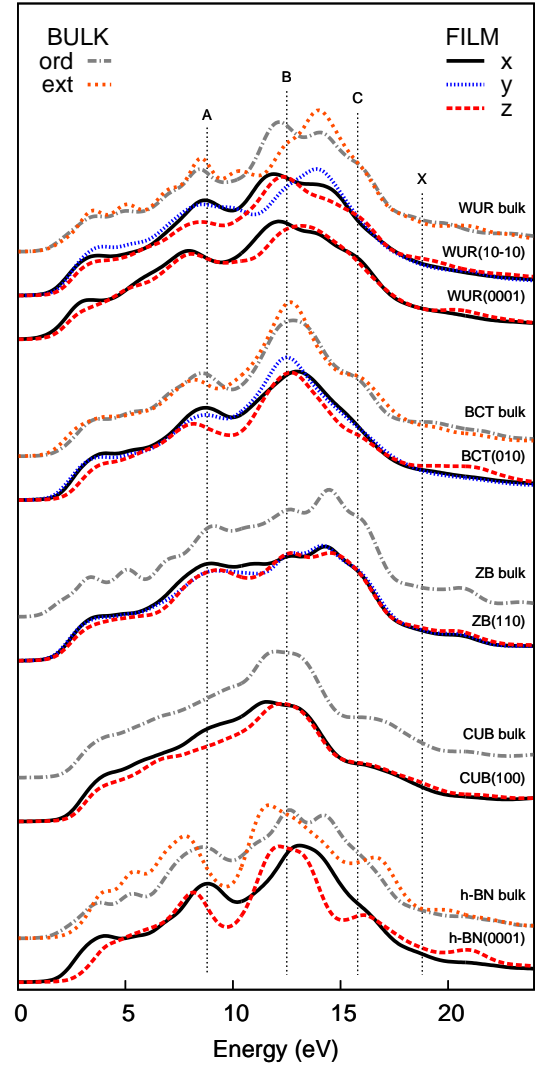}
	\caption{Optical absorption spectra of ZnO 4ML thin-films and bulks.
For the bulks, dashed-dotted black and dotted red lines mark the ordinary and extraordinary spectra.	In the films, solid black, dotted blue and dashed red lines are for the $\textbf{q}\parallel x$, $\textbf{q}\parallel y$,  $\textbf{q}\parallel z$ spectra, respectively. 
	CUB(100) and $h$-BN(0001) are isotropic on the $xy$ plane, so only the $\textbf{q}\parallel x$ component is reported.
	Dotted vertical lines arbitrarily aligned with structures of WUR(10$\bar{1}$0), highlight A, B, C, and X regions of the spectra. All spectra have been convoluted with a 0.5 eV wide Gaussian function. }
	\label{fig:abs_4ML}
\end{center}
\end{figure}

\subsection{Absorption spectra}

The absorption spectra (frequency dependence of the imaginary part of the dielectric function in the limit of vanishing momentum transfer) of the six ZnO thin films are displayed in Figure \ref{fig:abs_4ML}, together with their bulk reference. As for the high frequency dielectric constant, the anisotropy of the films is reflected in an increase of nonequivalent contributions in the absorption spectra.

Beyond the onset region to be discussed below, the film absorption spectra present many similarities with their corresponding bulks. The main absorption peaks, labeled A, B  and the  shoulder C correspond  to transitions from the upper part of the VB (from 0 to $\approx$ 3 eV below the VB maximum),  to the lower part of the CB (from 0 to $\approx$ 4 eV above the CB minimum), to the S/P$_1$ region, and to the P$_2$ region, respectively, with reference to the LDOS structures (Figure \ref{fig:dos}). Transitions from the Zn {$3d$} states have a small weight and their contributions to this part of the spectrum is negligible. At variance, the small X  structure located around $\omega = 20$ eV in both OC and EC is due to transitions from these localized Zn {$3d$} states to the P$_2$ region. Due to the differences in local environments of the I, $U_1$, $U_2$ and $U_3$ atoms and the LDOS structure shifts which result, the absorption peaks in the films are generally broadened with respect to their bulk counterparts.

\begin{figure}[h]
	\centering
	\includegraphics[height=0.7\textheight]{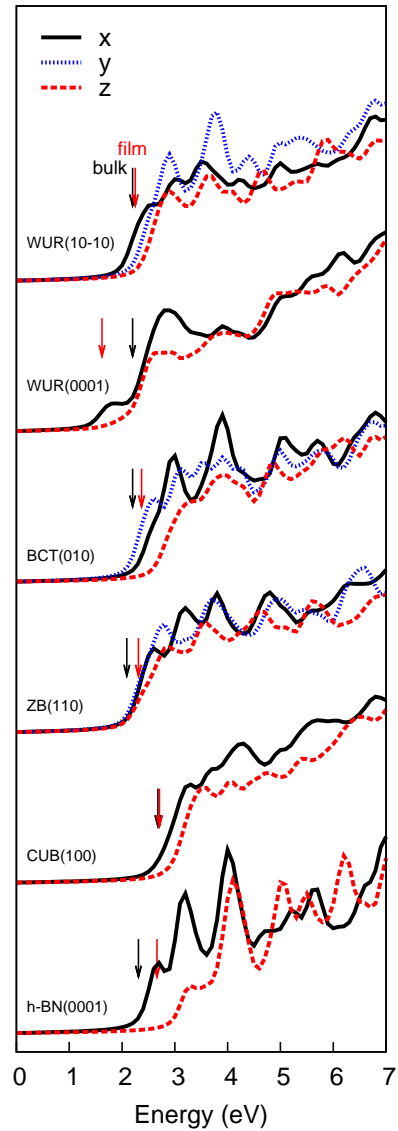}
\caption{Optical absorption spectra of ZnO 4ML thin-films close to threshold. Spectra have been convoluted with a 0.1 eV wide Gaussian function. The red and black arrows mark the absorption threshold in the films and their respective bulks, respectively. }
	\label{fig:abs_4ML_th}
\end{figure}

The main differences between the bulk and film absorption spectra occur in the vicinity of the absorption threshold, as seen in Figure \ref{fig:abs_4ML_th} which displays an enlarged view of this spectral region. First, the threshold energy is slightly shifted with respect to the bulk due to the gap variations (cf Table \ref{tab:dos_peaks}). In most polymorphs, this shift is very small and hardly visible at the scale of the figure. Only in WUR(0001) can it be well observed, due to the larger reduction of Madelung potential on surface atoms.

Specific structures are observed in the $\omega$ range approximately 0-3 eV above threshold, which involve transitions between  surface states/resonances in close vicinity to the VB maximum and CB minimum. Starting from the threshold, the first peaks which appear are in the $x$ and $y$ components of the spectra, consistently with the mostly $p_x$ and $p_y$ orbital character of the surface states at the top of the valence band. In the WUR(0001) absorption spectrum, a well-defined pre-peak exists due to a transition between the surface states at the top of the VB and the bottom of the CB.  

\begin{table}[h]
	\centering
	\caption{Correlation between the degree of hybridization of $p_z$ and $p_x,p_y$ orbitals at the top of VB  (measured by the dangling bond angle $\theta$), the energy difference $\Delta$ between the $z$ and $x,y$ first absorption peaks  and the anisotropy of the optical dielectric constant $\epsilon _\infty ^{x,y}-\epsilon _\infty ^{z}$.}
	\begin{tabular}{| l | c| c| c | }
	\hline
	&  $\theta$ ($^\circ$)& $\Delta$ (eV) & $\epsilon _\infty ^{x,y}-\epsilon _\infty ^{z}$\\
	\hline
h-BN(0001)&0&-0.6&0.44 \\
BCT(010)&6&-0.6&0.40\\
WUR(10$\bar{1}$0)&10&-0.4&0.4,0.32\\
CUB(100)&12&-0.4&0.27\\
ZB(110)&28&-0.2&0.24,0.16\\
	\hline	
	\end{tabular}
	\label{tab:anis}
	\end{table}

The threshold and first peaks in the $z$ absorption spectra occur at slightly higher energies than in the $x$ and $y$ spectra. Except in WUR(0001) in which electrostatic effects play a prominent role, this energy difference $\Delta$  inversely correlates with the degree of hybridization of the $p_z$ orbitals with the $p_x$ and $p_y$ ones, discussed in Section \ref{sec:electronic-DOS}. It decreases in the series h-BN(0001), BCT(010), WUR(10$\bar{1}$0), CUB(100), ZB(110), which is consistent with the decrease of anisotropy of the optical dielectric constant displayed in Table \ref{tab:cstopt}.

\section{Conclusion}
The present work, which relies on gradient-corrected and hybrid first principles simulations, provides a complete characterization of the electronic properties of ZnO thin-films with four monolayer  thickness, cut along the  BCT(010), CUB(100), h-BN(0001),  ZB(110), WUR(10$\overline{1}$0) and (0001) orientations. 

The modifications of the local densities of states have been described and analyzed in terms of the reduction of the Madelung potential on under-coordinated atoms and surface states/resonances appearing at the top of the VB and bottom of the CB. The gap width in the films is found to be larger than in the corresponding bulk, which is assigned to quantum confinement effects.

The components of the high frequency dielectric constant have been determined. They are larger than their bulk counterparts and display a larger anisotropy. Finally, the absorption spectra of the films have been computed. They display specific features in an energy range just above threshold due to transition from or to surface states/resonances.

This study provides a first understanding of finite size effects on the electronic properties of ZnO thin films. Analysis of their thickness dependence is currently under progress in our group.

\section*{Acknowledgments}
We  gratefully acknowledge  generous allocations of computing time at at GENCI- [TGCC/CINES/IDRIS] under project 100170.


\widetext

\section*{Supplemental Material}

This document provides  detailed information on the ZnO thin-films with four monolayer (4ML) thickness, cut along the  BCT(010), CUB(100), h-BN(0001),  ZB(110), WUR(10$\overline{1}$0) and (0001) orientations. In Table \ref{tab:distances2}, the Zn-O bond lengths  around nonequivalent Zn or O atoms are given. Figure \ref{fig:DOS} shows an enlarged view of the LDOS at the top of the valence band, projected on oxygen U$_1$, U$_2$, I, and in the case of WUR(0001) U$_3$ atoms, with $p_x$, $p_y$ or $p_z$ character.

\onecolumngrid

\begin{figure}[b!]
 	\includegraphics[width=0.33\textwidth]{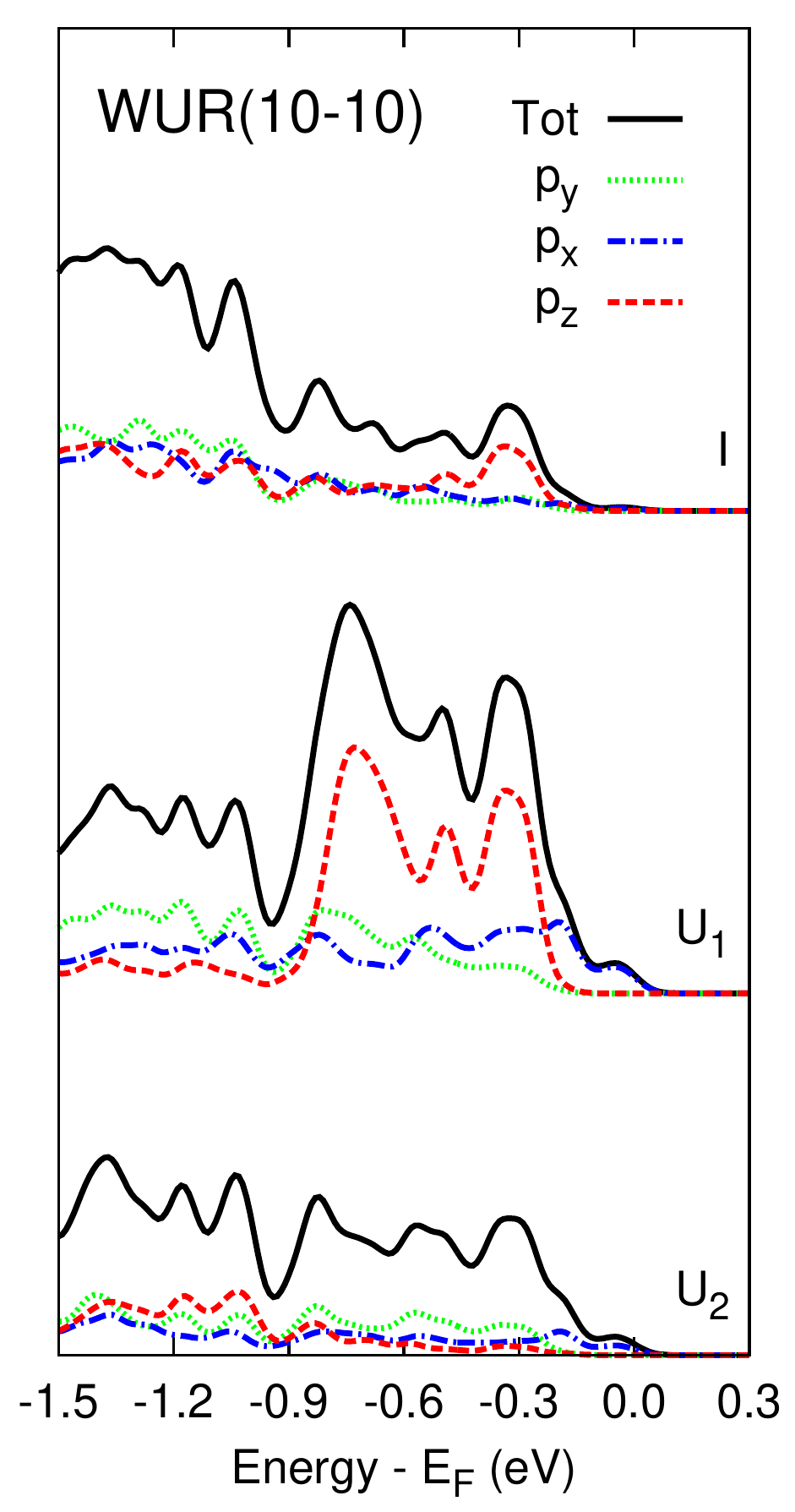}
 	\qquad \qquad
 	\includegraphics[width=0.33\textwidth]{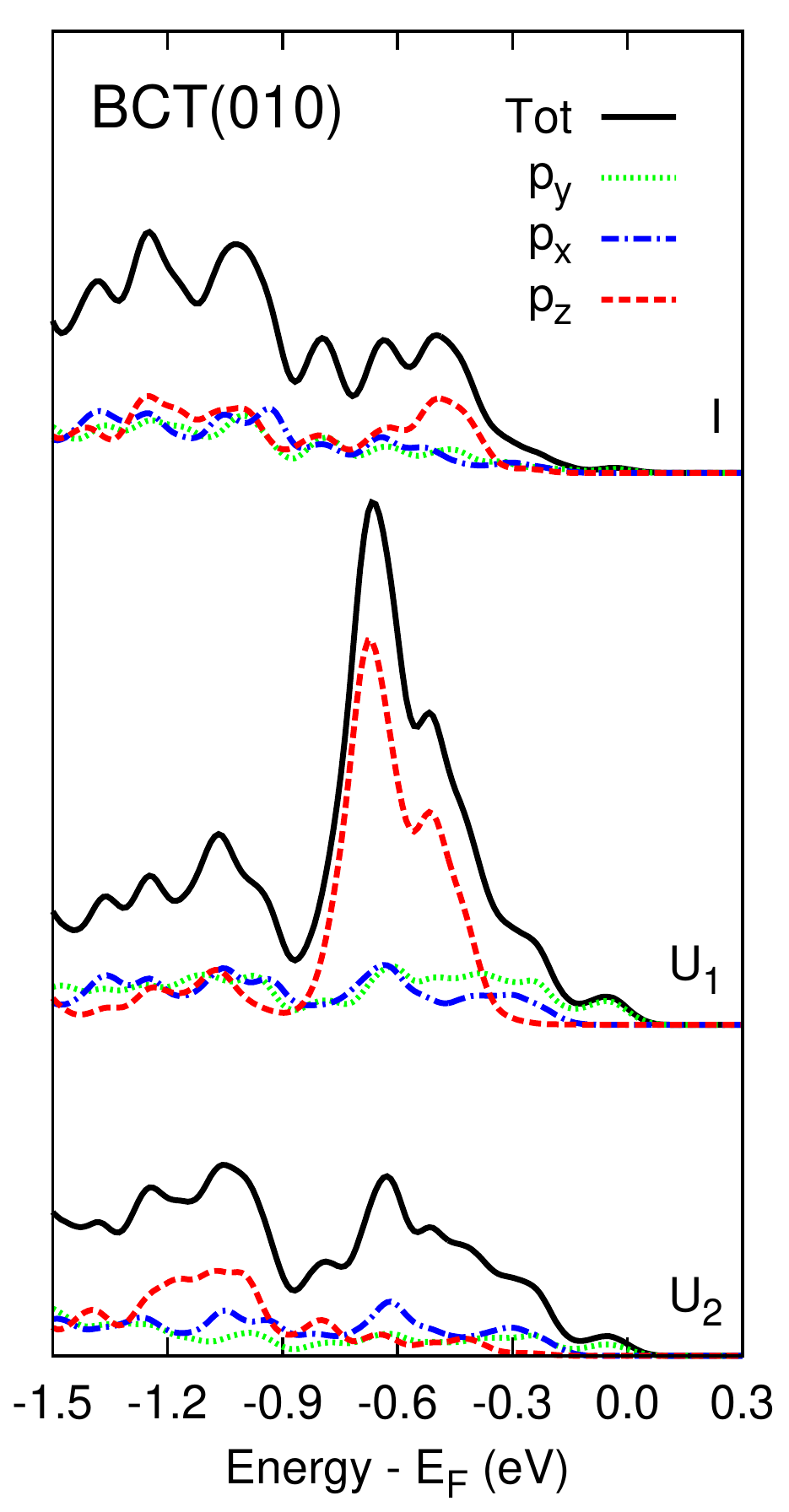}
\end{figure}
\twocolumngrid

\pagebreak
\begin{figure*} 
 	\includegraphics[width=0.33\textwidth]{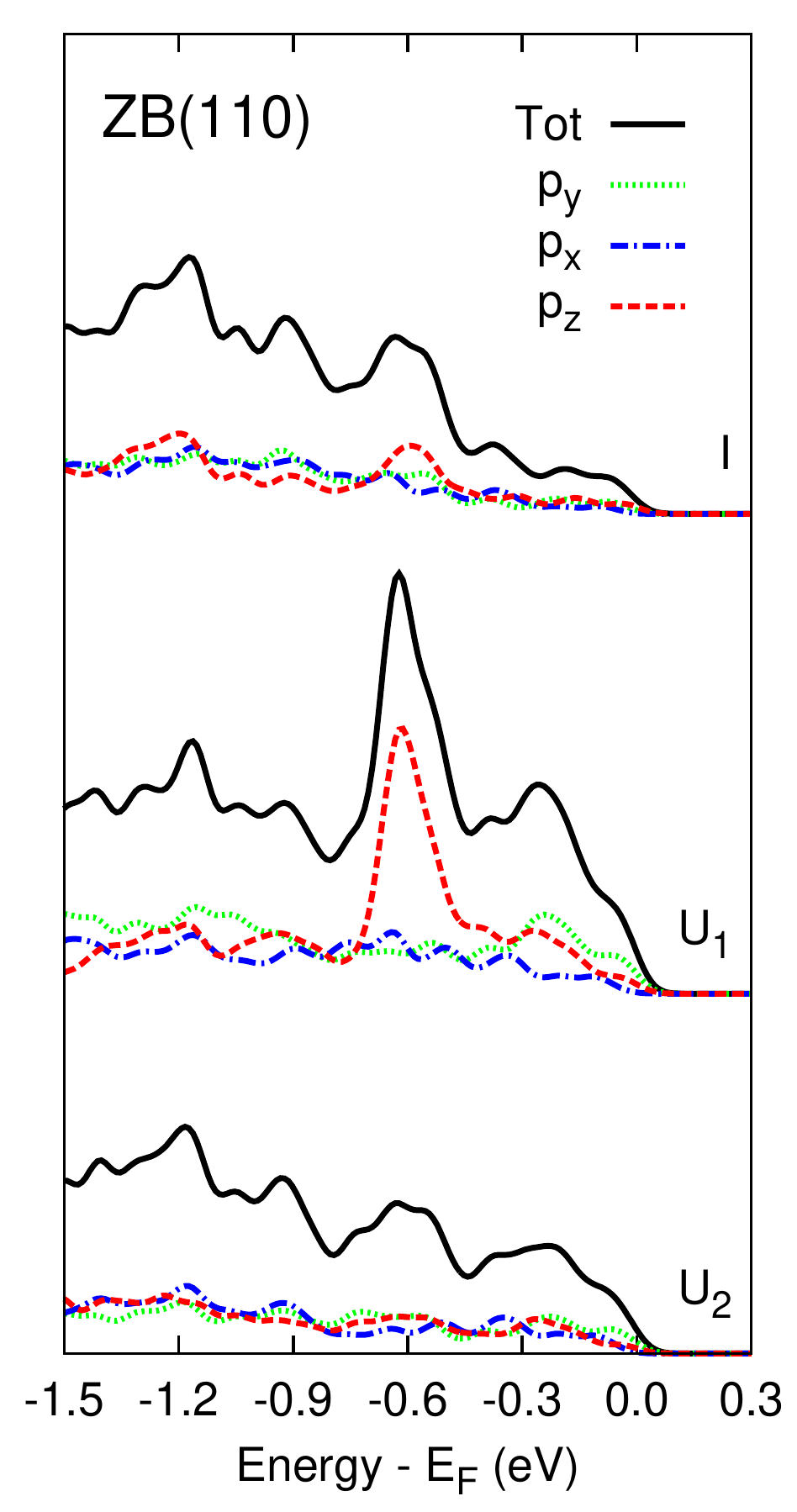}
 	\qquad	\qquad
 	\includegraphics[width=0.33\textwidth]{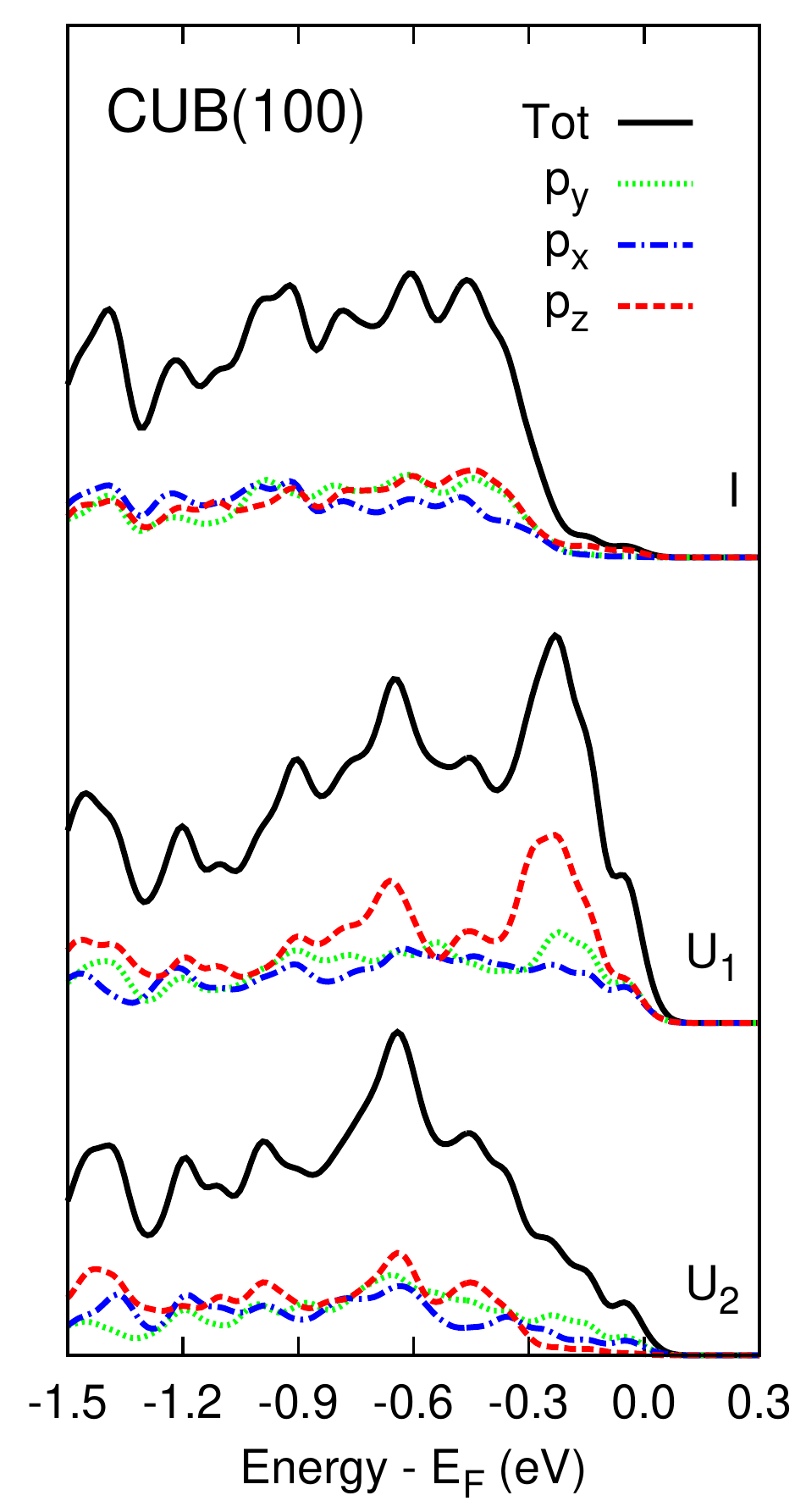}
 	
 	\vspace{1cm}

 	\includegraphics[width=0.33\textwidth]{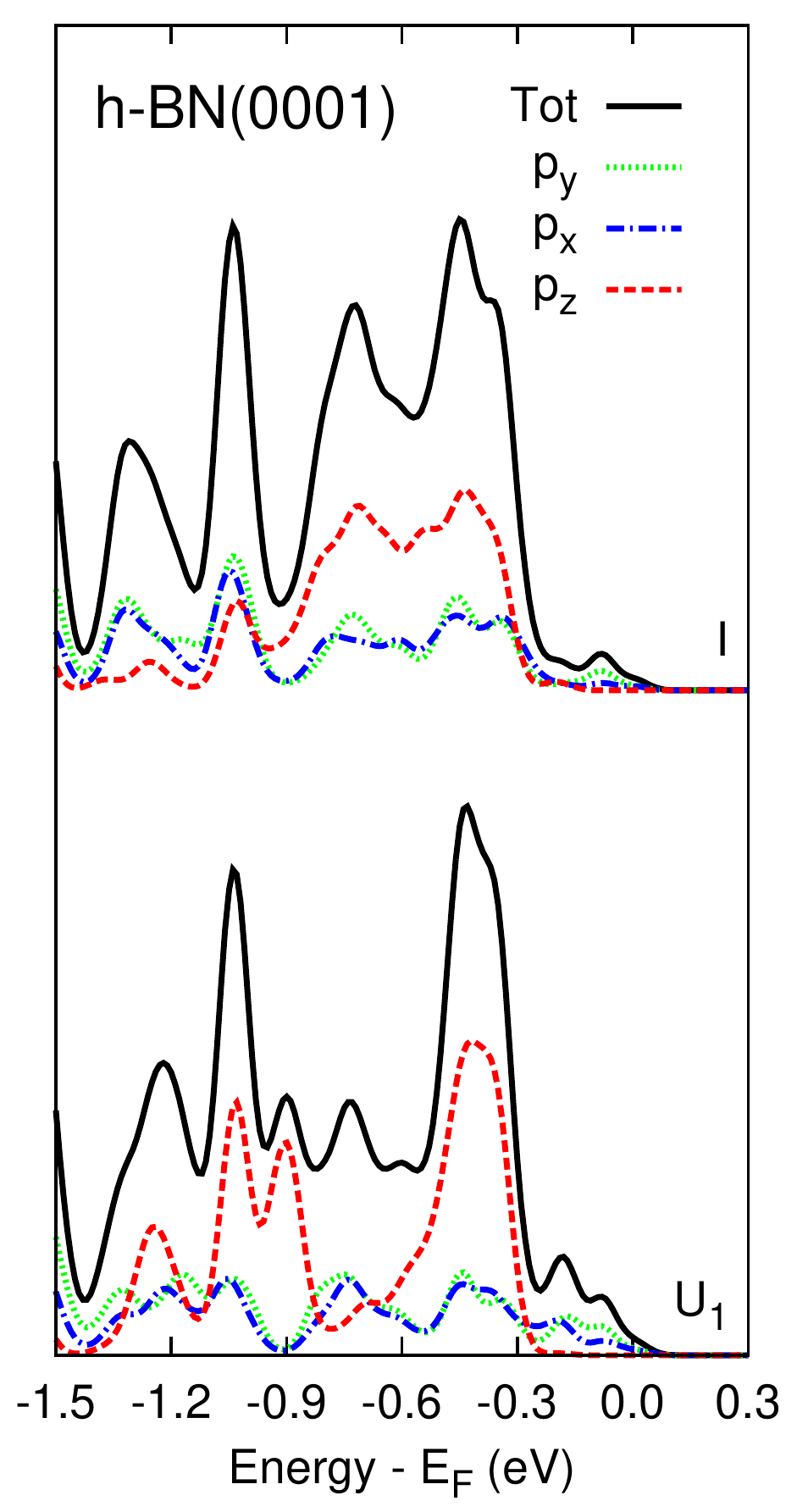}
 	\qquad	\qquad
 	\includegraphics[width=0.33\textwidth]{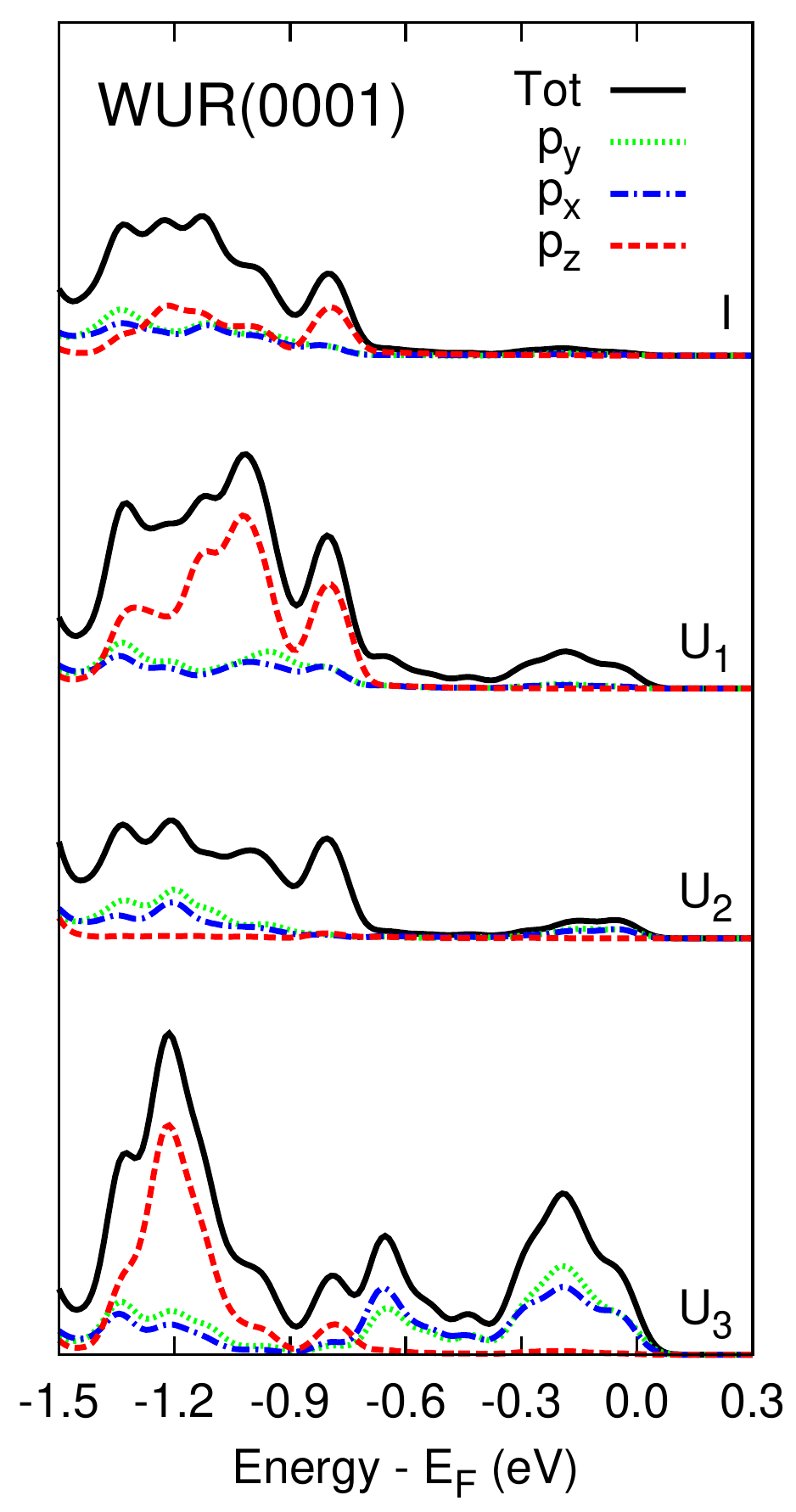}
 	\caption{(Colors online:) Local densities of states in the 4ML films on I, $U_1$, $U_2$ and $U_3$ (in the case of WUR(0001)) atoms, represented as solid black,  dashed-dotted blue,  dotted-green, and dashed red lines, respectively, with specification of the $p_x$, $p_y$ or $p_z$ character.}
 	\label{fig:DOS}
\end{figure*}


\begin{table*}
	\centering
   \caption{Zn-O bond lengths (\AA) around nonequivalent Zn (left part) or O (right part) atoms in the ZnO 4ML polymorph films.}
	\begin{tabular}{| l   c c c c  |   c c c   c |}
	\hline
			& \multicolumn{4}{ c}{Zn} & \multicolumn{4}{|c|}{O} \\
	\hline
	Wur(10$\bar{1}$0): & & & & & & & & \\
	\quad 	U$_1$     &1.87&1.95&1.95&- &1.87&1.94&1.94&-  \\ 
	\quad 	U$_2$    &1.94&1.94&2.01&2.11 &1.95&1.95&2.01&2.08   \\
	\quad 	I & 1.99&1.99&2.00&2.08&1.99&1.99&2.00&2.11    \\
	         & 1.99&1.99&2.02&2.03&1.99&1.99&2.02&2.03\\
	\quad {\it bulk} & {\it 2.00} & \multicolumn{3}{c|}{-} & {\it 2.00} & \multicolumn{3}{c|}{ - } \\
	\hline
		Bct(010):  & & & & & & & & \\
	\quad 	U$_1$     &1.87&1.93&1.93&- &1.87&1.95&1.95& - \\ 
	\quad 	U$_2$    &1.95&1.95&2.04&2.12 &1.93&1.93&2.04&2.14   \\
	\quad 	I & 1.97&1.97&2.03&2.08&1.97&1.97&1.98&2.12    \\
	         & 1.96&1.96&1.98&2.14&1.96&1.96&2.03&2.08\\
	\quad {\it bulk} & {\it 2.00} & \multicolumn{3}{c|}{-} & {\it 2.00} & \multicolumn{3}{c|}{ - } \\
	\hline
		Zb(110):  & & & & & & & & \\
	\quad 	U$_1$     &1.88&1.88&1.98&- &1.88&1.88&1.99&-  \\ 
	\quad 	U$_2$    &1.99&1.99&1.99&2.03 &1.98&1.99&1.99&2.03   \\
	\quad 	I & 2.00&2.01&2.01&2.03&2.00&2.01&2.01&2.03    \\
	         & 2.00&2.01&2.01&2.01&2.00&2.01&2.01&2.01  \\
	\quad {\it bulk} & {\it 2.00} & \multicolumn{3}{c|}{-} & {\it 2.00} & \multicolumn{3}{c|}{ - } \\
	\hline
		Cub(100): & & & & & & & &  \\
	\quad 	U$_1$     &1.86&1.93&1.93&- &1.86&1.93&1.93&- \\ 
	\quad 	U$_2$    &1.86&1.86&1.86&2.10 &1.86&1.86&1.86&2.09   \\
	\quad 	I & 1.88&2.03&2.03&2.09&1.88&2.03&2.03&2.10   \\
	         & 1.88&1.93&1.93&2.06&1.88&1.93&1.93&2.06\\
	\quad {\it bulk} & {\it 2.01} & \multicolumn{3}{c|}{-} & {\it 2.01} & \multicolumn{3}{c|}{ - } \\
	\hline
		$h$-BN(0001): & & & & & & & &  \\
	\quad 	U$_1$     &1.96&1.96&1.96&2.32 &1.96&1.96&1.96&2.36 \\ 
	\quad 	I & 1.96&1.96&1.96&2.36$\times$2&1.96&1.96&1.96&2.35$\times$2  \\
	\quad {\it bulk} & {\it 2.09$\times$3} &  {\it 2.30$\times$2}&\multicolumn{2}{c|}{-} & {\it 2.09$\times$3} &  {\it 2.30$\times$2}&\multicolumn{2}{c|}{-}\\
	\hline
		Wur(0001)-Zn: & & & & & & & & \\
	\quad 	U$_1$     &1.89&1.89&1.97&- &- &- & -& - \\ 
	\quad 	U$_3$     &- &- &- &- &1.89&1.89&1.97& - \\ 
	\quad 	U$_2$    &-&- &- & - &1.97&1.97&1.97&2.12  \\
	\quad 	I & 1.97&2.00&2.05&2.05&1.98&2.00&2.00&2.00    \\
	         & 1.99&1.99&1.99&2.12&1.99&2.00&2.05&2.05\\
	\quad {\it bulk} & {\it 2.00} & \multicolumn{3}{c|}{-} & {\it 2.00} & \multicolumn{3}{c|}{ - } \\
	\hline
		Wur(0001)-0: & & & & & & & & \\
	\quad 	U$_1$     &- &- &- &- &1.89&1.89&1.95& - \\ 
	\quad 	U$_3$     &1.89&1.89&1.99&- &- &- &- & - \\ 
	\quad 	U$_2$   &1.95&1.95&1.95&2.09 &- &- &- & -   \\
	\quad 	I & 1.99&2.00&2.04&2.04&1.99&1.99&1.99&2.09    \\
	         & 1.98&2.00&2.00&2.00&1.99&2.00&2.04&2.04\\
	\quad {\it bulk} & {\it 2.00} & \multicolumn{3}{c|}{-} & {\it 2.00} & \multicolumn{3}{c|}{ - } \\
	\hline	
	\end{tabular}
	\label{tab:distances2}
\end{table*}


\begin{thebibliography}{99}
\bibitem{noteelectrostatic}
 Average electrostatic potentials are evaluated by locating a test charge with norm unity at each ion core, as implemented in VASP\cite{VASP_code}.
\bibitem{anisotropy}The anisotropy $\epsilon_\infty^x-\epsilon_\infty^y$ in the BCT(010) film is less than 0.01 and thus not visible in Table \ref{tab:cstopt}.
\end{thebibliography}
\end{document}